%% file: main.tex
\documentclass[conference]{IEEEtran}
\IEEEoverridecommandlockouts
\usepackage{cite}
\usepackage{amsmath,amssymb,amsfonts}
\usepackage{algorithm,algorithmicx}
\usepackage{algpseudocode}

\usepackage{amsmath}  
\usepackage{caption}
\usepackage{subcaption}
\usepackage{graphicx}
\usepackage{textcomp}

\usepackage{flushend}

\usepackage{dcolumn}
\usepackage{bm}
\usepackage{braket}
\usepackage{hyperref}
\usepackage{mwe}
\usepackage{comment}

\usepackage{float}
\usepackage[section]{placeins}

\hypersetup{
    colorlinks = true,
    citecolor = magenta,
    linkcolor = purple
}
\def\BibTeX{{\rm B\kern-.05em{\sc i\kern-.025em b}\kern-.08em
    T\kern-.1667em\lower.7ex\hbox{E}\kern-.125emX}}

\input{custom-command}

\usepackage{siunitx}    
\usepackage{ragged2e}
\usepackage{booktabs, makecell, multirow, tabularx}
\setcellgapes{2pt}
\newcolumntype{L}[1]{>{\RaggedRight\hspace{0pt}%
                     \hsize=#1\hsize}X}
\usepackage{lipsum}

\begin{document}
\bstctlcite{IEEEexample:BSTcontrol}

\title{A Substrate Scheduler for Compiling Arbitrary Fault-tolerant Graph States 

\thanks{This research was developed in part with funding from the Defense Advanced Research Projects Agency [under the Quantum Benchmarking (QB) program under award no. HR00112230007 and HR001121S0026 contracts]}
\thanks{This work was supported by MEXT-Quantum Leap Flagship Program Grant Number JPMXS0118067285, JPMXS0120319794.}
\thanks{DM acknowledges support from the Sydney Quantum Academy.}}

\author{
\IEEEauthorblockN{
    Sitong Liu\IEEEauthorrefmark{1}\IEEEauthorrefmark{4},
    Naphan Benchasattabuse\IEEEauthorrefmark{1}\IEEEauthorrefmark{4},
    Darcy QC Morgan\IEEEauthorrefmark{5},\\
    Michal Hajdu\v{s}ek\IEEEauthorrefmark{1}\IEEEauthorrefmark{4},
    Simon J. Devitt\IEEEauthorrefmark{5},
    and Rodney Van Meter\IEEEauthorrefmark{3}\IEEEauthorrefmark{4}
}\\
\IEEEauthorblockA{\IEEEauthorrefmark{1}\textit{Graduate School of Media and Governance, Keio University Shonan Fujisawa Campus}, Kanagawa, Japan}

\IEEEauthorblockA{\IEEEauthorrefmark{3}\textit{Faculty of Environment and Information Studies, Keio University Shonan Fujisawa Campus}, Kanagawa, Japan}

\IEEEauthorblockA{\IEEEauthorrefmark{4}\textit{Quantum Computing Center, Keio University}, Kanagawa, Japan}

\IEEEauthorblockA{\IEEEauthorrefmark{5}\textit{Centre for Quantum Software and Information, University of Technology Sydney}, Sydney, NSW 2007, Australia\\
\{sitong,whit3z,michal,rdv\}@sfc.wide.ad.jp, 
darcy.qc.morgan@gmail.com, simon.devitt@uts.edu.au}

}

\maketitle
\thispagestyle{plain}
\pagestyle{plain}

\begin{abstract}
\input{abstract}
\end{abstract}


\section{Introduction} \label{Introduction}
\input{introduction}

\section{Preliminaries} \label{Preliminaries}
\input{preliminaries}

\section{Our Approach} \label{Approach}
\input{ourapproach}

\section{Evaluation} \label{Evaluation}
\input{evaluation}

\section{Discussion} \label{Discussion}
\input{discussion}

\appendices
\section{Pseudocode for MinCut Mapper}
\input{Appendix}

\label{Appendix}

\section*{Code Availability}
Substrate Scheduler’s source code, documentation, and sample
configurations are fully available online \footnote{\url{https://github.com/sfc-aqua/gosc-graph-state-generation}}. In addition to the tool, we also provide benchmarking code for several sample runtime experiments, along with a visualization tool. Feedback and requests for features are welcome. 

\bibliographystyle{IEEEtran.bst}
\bibliography{IEEEabrv, bibfile}

\end{document}

%% file: custom-command.tex
\usepackage[svgnames, table]{xcolor}
\usepackage{tabularx, makecell, linegoal}

\usepackage{ifthen}

\newboolean{ShowComments}
\setboolean{ShowComments}{true}  
\ifthenelse{\boolean{ShowComments}}%
	{
		\newcommand{\ColorComment}[3]{%
				{\colorbox{#1}{\color{White}   \textsf{\textbf{#2}}} \textcolor{#1}{#3}}}

	}%
	{
		\newcommand{\ColorComment}[3]{}

	}%

\definecolor{rdvcolor}{rgb}{0,0.5,0}
\definecolor{michalcolor}{RGB}{255,127,80}
\definecolor{naphancolor}{RGB}{165,6,245}
\definecolor{sitongcolor}{rgb}{0,0.5,0.8}
\definecolor{simoncolor}{rgb}{0,0.5,0}
\definecolor{darcycolor}{RGB}{254,0,0}



%% file: abstract.tex
Graph states are useful computational resources in quantum computing, particularly in measurement-based quantum computing models.
However, compiling arbitrary graph states into executable form for fault-tolerant surface code execution and accurately estimating the compilation cost and the run-time resource cost remains an open problem.
We introduce the Substrate Scheduler, a compiler module designed for fault-tolerant graph state compilation. 
The Substrate Scheduler aims to minimize the space-time volume cost of generating graph states.
We show that Substrate Scheduler can efficiently compile graph states with thousands of vertices for ``A Game of Surface Codes''-style patch-based surface code systems. 
The results show that our module generates graph states with the lowest execution time complexity to date, achieving graph state generation time complexity that is at or below linear in the number of vertices and demonstrating specific types of graphs to have constant generation time complexity.
Moreover, it provides a solid foundation for developing compilers that can handle a larger number of vertices, up to the millions or billions needed to accommodate a wide range of post-classical quantum computing applications.

%% file: introduction.tex
Spanning almost half a century, the development of quantum computing, which leverages the principles of quantum mechanics, has progressed dramatically~\cite{ladd2010quantum}. Quantum computing has the potential to solve some problems that are either impossible or extremely difficult to solve with classical computers. However, an inevitable problem with quantum computers is that they are still small and highly susceptible to noise~\cite{preskill2018quantum}. 
In order to fully realize the potential of quantum computing and achieve the so-called ``quantum advantage'', it is of vital importance to guarantee the high-fidelity execution of large-scale quantum programs. In the long term, the pressing challenge is how to transition from the current generation of noisy quantum devices to fault-tolerant quantum computing with quantum error correction (QEC).

The number and fidelity of qubits, both physical and logical, will remain a constraint on applications for the foreseeable future.
In the meantime, significant resources must be allocated to error correction, as implementing quantum error correction involves using large numbers of physical qubits to encode a single fault-tolerant logical qubit~\cite{gottesman2010introduction,devitt13:rpp-qec,RevModPhys.87.307}.
As such, in order to hasten the advent of the scalable, fault-tolerant quantum computing era~\cite{preskill2018quantum}, efficient compilers that can implement applications using quantum error correction with minimal space-time cost are critical.

In recent years, engineers have gradually started to focus on how to construct and optimize fault-tolerant quantum computers and the related quantum error correction protocols~\cite{chamberland2020very}.
One problem that needs to be solved is that current research is focused either on the logical level, assuming all operations are already fault-tolerant, or on constructing fault-tolerant qubits~\cite{satzinger2021realizing}, i.e., how to encode and decode them. However, there is no integrated description to guide us on how to run programs on a fault-tolerant quantum computer.

\begin{figure*}[htb]
\centering
\includegraphics[width=0.9\textwidth]{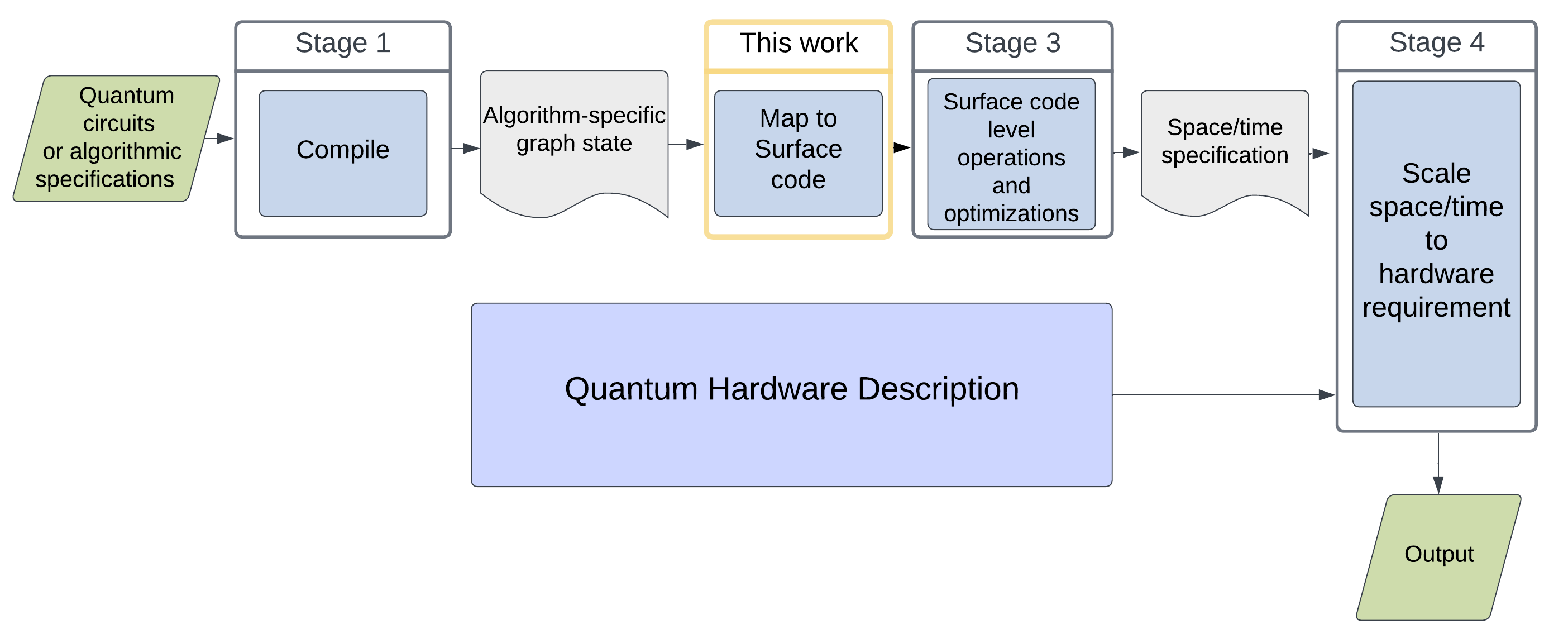}
\caption{The workflow for generating fault-tolerant graph states. Stage 1 converts the quantum circuit as input into a circuit/algorithm-specific graph state (i.e., the work of Jabalizer~\cite{algorithm-specific}). The second stage, which is the focus of this paper, involves compiling the graph state into a set of operations that can be executed on the surface code base. Stage 3 involves operations and optimization at the surface code level, and after final integration with the hardware, operations can be executed to generate the graph state.}
\label{Framea}
\end{figure*}

Recent work has demonstrated that arbitrary quantum circuits can be compiled into graph states that are believed to be amenable to further optimization and efficient execution~\cite{algorithm-specific}.
As a generalization of cluster states, graph states~\cite{hein2004multiparty,hein2006entanglement} have a variety of applications in quantum information, in particular as algorithmic resources in the context of measurement-based quantum computing (MBQC)~\cite{PhysRevLett.86.5188, NIELSEN2006147}. Recognizing their importance and flexibility, a group of researchers has proposed an end-to-end compilation toolchain based around the concept of fault-tolerant graph states, named {\bf benchq}~\footnote{\url{https://github.com/zapatacomputing/benchq}}, which consists of four stages (see Fig.~\ref{Framea}). The first stage, as studied by Vijayan \emph{et al.}, involves compiling quantum circuits/algorithms into graph states~\cite{algorithm-specific}. The main focus of this paper is on bridging the gap in Stage 2 of the toolchain, addressing how to map algorithmic graph states onto the surface code, a crucial step in achieving fault-tolerance.

In this paper, we present our implementation of Stage 2, the Substrate Scheduler, a compiler module that performs a fault-tolerant compilation of a graph state from the adjacency matrix to an optimized schedule of stabilizer measurement along with the logical qubit allocation. 
The Substrate Scheduler is based on the surface code~\cite{fowler-defect-surface-code, kitaev-surface-code, Dennis_2002}, one of the most promising quantum error correction codes, with lattice surgery~\cite{horsman-patch-surface-code}, using the rules introduced by Litinski’s paper, ``A Game of Surface Codes'', called GoSC~\cite{litinskiGameSurfaceCodes2019} (see Section \ref{Preliminaries}).

GoSC uses a space-time resource model for evaluating performance. The primary space-time trade-off we observe is between the number of logical qubits (measured in units of ``tiles'') and the code cycles (known as ``Tocks'') spent in generating the graph state.
Similar evaluations involving resource overhead related to time and space have been applied in various quantum computing studies~\cite{steane02:ft-qec-overhead, PhysRevA.100.032328, Paler_2017}. Therefore, this work also adopts the same overhead evaluation metric. The primary goal of the Substrate Scheduler is to minimize the space-time volume cost. We analyzed and optimized the layout design used to generate the graph state.
The stabilizer formalism~\cite{gottesman1997stabilizer} (see Section \ref{Preliminaries}), which was originally developed for the analysis and design of quantum error correction codes, is used to further optimize the generation process.  Our module consists of three parts: stabilizer generator reduction, heuristic algorithms for mapping graph vertices to logical qubits of surface code, and scheduling of the stabilizer generator measurements.

In summary, this paper makes the following contributions:
\begin{itemize}
    \item We investigate two methods of generating fault-tolerant quantum graph states based on surface codes and find that the minimum time step cost required for using stabilizer parity checks on the surface code is lower than that of preparing the graph state with CZ gates. We provide a demonstration of the required time and space costs based on the two-tile patch one-bus layout and analyze the relationship between space and time.
    \item We developed optimization techniques for generating graph states via the stabilizer formalism, and show that they can be used to reduce time costs.
    \item Our compiler module optimizes the process of synthesizing fault-tolerant graph states and allows for generating graph states with the lowest generation time complexity to date, achieving graph state generation time complexity that is at or below linear in the number of vertices and enabling specific types of graphs to have constant creation time complexity.
\end{itemize}

The rest of the paper is organized as follows: Section \ref{Preliminaries} provides an introduction to the key concepts of graph states, stabilizer formalism, and basic surface code operations that are utilized in this study. 
Section~\ref{Approach} provides an overview of the framework for the Substrate Scheduler and demonstrates the compiling process in detail. 
This includes pre-mapping optimization, scheduling of stabilizer measurement, and heuristic vertex-to-qubit mapping techniques. In Section \ref{Evaluation}, we present the results obtained from experiments conducted using the Substrate Scheduler. 
Finally, in Section \ref{Discussion}, we discuss future work and conclude.

%% file: preliminaries.tex
In this section, we give a brief overview of the required concepts and notation used throughout this manuscript.

\subsection{Graph State and Stabilizer Formalism}

We begin with the notion of a \emph{graph} $G = (V, E)$~\cite{west2001introduction}, which is a pair of two sets, a vertex set $V=\{ 1, 2, 3, \ldots, n \}$ and an edge set $E \subset  V^2$.
We also denote $|V| = n$ and $|E|$ as the number of vertices and the total number of edges, respectively.
Two vertices $a, b\in V$ are \emph{adjacent} if they are connected by an edge, $\{a,b\}\in E$.
In this work, we only consider \emph{connected simple graphs} where only at most one edge is allowed per pair of vertices with no self-loops and there exists a path between any pair of vertices.
This gives rise to the notion of an \emph{adjacency matrix} $\Gamma = [\Gamma_{a, b}] \in \{0, 1\}^{n \times n}$ with elements
\begin{equation}
    \Gamma_{a,b} = \begin{cases} 1, & \text{if } \{a,b\}\in E \\
    0, & \text{otherwise}. \end{cases}
\end{equation}
We also make repeated use of the \emph{neighborhood} $\operatorname{ngbr}(a)$ of a vertex $a\in V$,
\begin{equation}
    \operatorname{ngbr}(a) = \{ b\in V | \{a,b\}\in E \}.
\end{equation}
The neighborhood is the set of vertices adjacent to a given vertex.
A subset of vertices $U \subset V$ is \emph{independent} if no two of its vertices are adjacent.
The \emph{maximum independent set}, $\alpha(G)$, is the largest such set.

Central to our discussion is the notion of \emph{graph states} \cite{hein2004multiparty,hein2006entanglement}, a particular class of multipartite entangled states.
The vertices correspond to the qubits in the system while edges correspond to interactions between pairs of qubits.
For a given graph $G$, we can construct its corresponding graph state $\ket{G}$ by first initializing all qubits in the state $|+\rangle=(|0\rangle+|1\rangle)/\sqrt{2}$, and for each pair of qubits which represent adjacent vertices in $G$, an entangling two-qubit \emph{controlled-phase gate},
\begin{equation}
    CZ = |0\rangle\langle 0| \otimes I + |1\rangle\langle 1| \otimes Z,
\end{equation}
is applied.

Since graph states are a subclass of more general stabilizer states~\cite{gottesman1997stabilizer}, there is a compact description of a state using only $n$ operators of length at most $n$, in contrast to the full state vector representation, which requires a complex amplitude for each of the $2^n$ basis states.
The stabilizer formalism describes quantum states in terms of operators that \emph{stabilize} the state.

A state $|\psi\rangle$ is stabilized by an operator $K$ if $K|\psi\rangle=|\psi\rangle$, that is, the state $|\psi\rangle$ is a +1 eigenstate of the operator $K$.
For example, state $|+\rangle$ is stabilized by the Pauli operator $X$ since $X|+\rangle=|+\rangle$.
We associate a multi-qubit Pauli operator with qubit $i$ of a graph state with the following form,
\begin{equation}
    g_i = X_i \bigotimes_{j\in \operatorname{ngbr}(i)} Z_j.
    \label{eq:stabilizer_generator}
\end{equation}
The $n$-qubit graph state $|G\rangle$ is then uniquely identified as the simultaneous +1 eigenstate of all $n$ stabilizer operators from Eq.~(\ref{eq:stabilizer_generator}).
These stabilizer operators generate an Abelian group referred to as the \emph{stabilizer} $\mathcal{S}=\langle g_1, \ldots, g_n\rangle$.
Stabilizer generators of Eq.~(\ref{eq:stabilizer_generator}) lead to an efficient description of the graph state in terms of $n$ commuting operators.

We will make repeated use of measuring the stabilizer generators.
Measurement of stabilizer generator $g_i$ corresponds to the application of a projection operator onto the even/odd parity subspace for the stabilizer generator,
\begin{equation}
    \Pi^{\pm}(g_i) = \frac{1}{2} (I \pm g_i).
\end{equation}
If the measured parity is even, that is, we have projected onto the positive eigenspace of the stabilizer generator, we do not need to take further action.
Projections onto the negative eigenspace can be further corrected to flip their parity, but in general, this is classically tracked without applying quantum gates during the generation process.

Once the generation of the graph state is complete, in the absence of decoherence, it is clear from our previous discussion that measurement of any of the stabilizer generators of the graph state $|G\rangle$ produces a $+1$ outcome with unit probability.
Similarly, projecting any initial state onto the common even parity eigenspace of all the stabilizer generators prepares the desired graph state $|G\rangle$.

\subsection{Surface Code}

We employ the surface code\cite{fowler-defect-surface-code, kitaev-surface-code, Dennis_2002}, which encodes logical qubit states into the collective state of a lattice of physical qubits, as the method for quantum error correction. The surface code's nearest neighbor interaction structure in a two-dimensional plane offers natural and practical realizability compared to non-local codes or codes that require transversal operations. Additionally, its high tolerance to errors \cite{threthold, Wang} makes it an ideal candidate for error correction in quantum computing.

There are many variations of surface codes, such as the defect-based~\cite{fowler-defect-surface-code}, the twist-based~\cite{bombin-twisted-surface-code}, or the patch-based~\cite{horsman-patch-surface-code} encodings.
We focus on the patch-based with the simplified rules of tile-based board games introduced in~\cite{litinskiGameSurfaceCodes2019}.
The board is partitioned into a number of tiles where they can host \emph{patches} representing qubits.
The basic rules of this tile game can be summarized as follows.

\subsubsection{Correspondence to surface code lattice surgery}
Assuming that we are using the surface code with a code distance $d$, a tile on the board represents $\sim 2d^2$ physical qubits.
Approximately half of these qubits are used for the data state, while the other half are used for error syndrome extraction.
The solid and dashed edges of a patch correspond to the logical $Z$ and $X$ operators on the boundaries of the surface code.
The performance metric of this tile-based game is the space-time volume which corresponds to the board area and the unit of time corresponding to the $d$ error-check code cycles.
To describe the time required for operations on patches, a unit of time called \emph{Tock} is introduced, which is exactly $d$ rounds of code cycles. 
0 Tock is a special case that does not mean 0 code cycles; rather, it represents operations that have a constant time cost and does not scale with the code distance $d$. However, it should be noted that the constant time cost may be non-zero.

\subsubsection{Logical qubit representation}

\begin{figure}[tbp]
    \centering
    \includegraphics[width=0.25\textwidth]{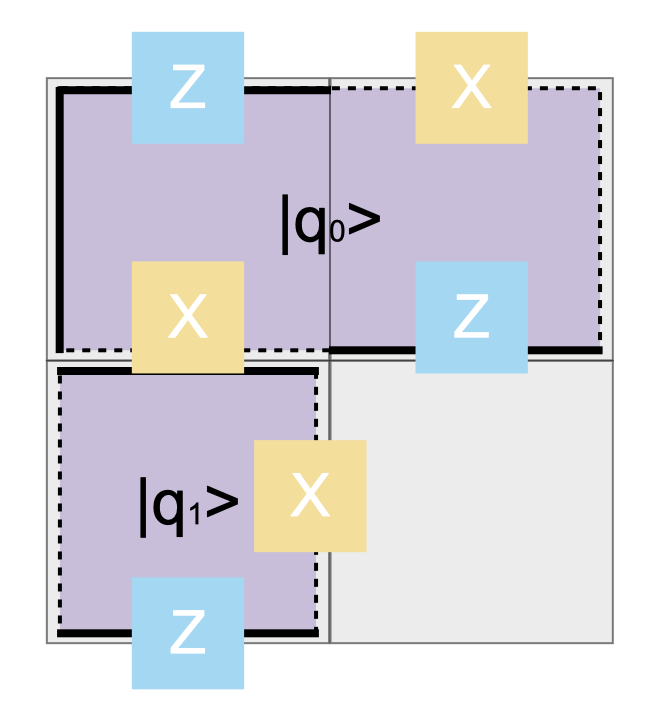}
    \caption{Examples of two different kinds of one-qubit patches in a $2\times 2$ grid of tiles. A patch (logical qubit) can occupy one or more tiles. Patches have dashed and solid edges representing Pauli operators. The dashed edges represent the qubit’s $X$ operator, solid edges represent the qubit’s $Z$ operator. }
    \label{fig:patch-examples}
\end{figure}

A \emph{patch} is a contiguous area, which can span over multiple tiles, used to represent one or more logical qubits (see Fig.~\ref{fig:patch-examples}). 
In this work, we will only use one logical qubit per patch.
On the boundary of the patch, there can be solid or dashed edges representing the Pauli $Z$ and $X$ operators, respectively. 
The point at which the solid and dashed edges meet is called a corner or $X$/$Z$ corner, even if the two edges are on the same line, for historical reasons. 
We will see their importance when we discuss operations on patches.

\subsubsection{Patch operations}

We will now describe the native operations available in this tile-based game with their corresponding time cost in units of Tocks.
 
\paragraph{Qubit initialization} One-qubit patches can be initialized in the +1 eigenstates of $X_L$ or $Z_L$ basis ($\ket{+}_L$ or $\ket{0}_L$) in $0$ Tocks, where subscript $L$ is used to denote that these are logical Pauli operations and states.

\paragraph{Single-patch measurements} One single patch can be measured in either the $X$ or $Z$ basis.
After the measurement, the patch will be removed from the board, freeing up previously occupied tiles. 
This operation has a cost of $0$ Tocks.

\paragraph{Multi-qubit Pauli product measurement/parity measurement}

\begin{figure}[tbp]
\centering
\includegraphics[width=0.5\textwidth]{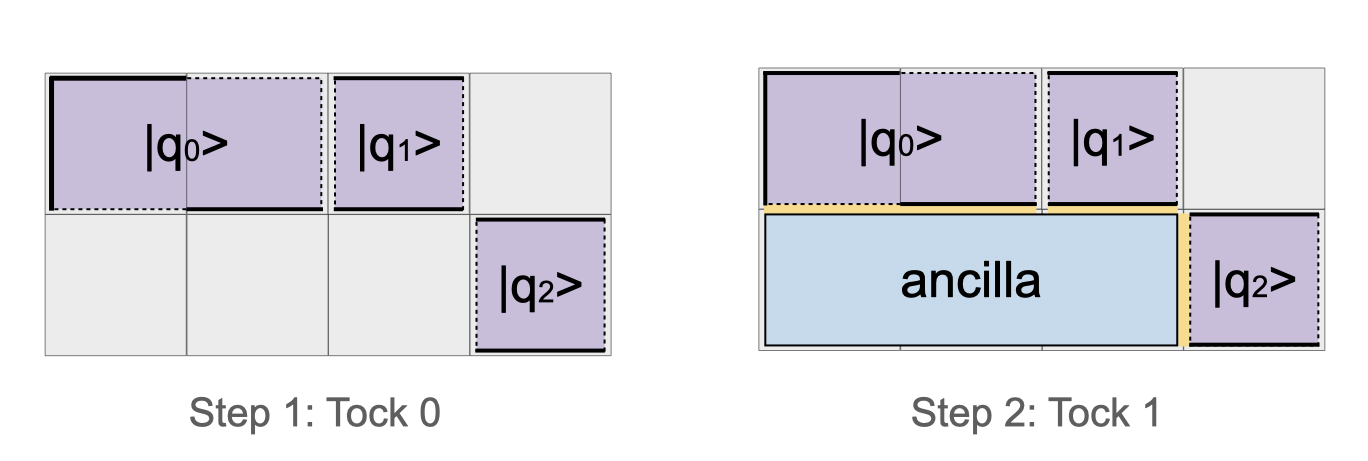}
\caption{
A multi-qubit $Y\ket{q_0} \otimes Z\ket{q_1}\otimes X\ket{q_2}$ measurement in one time step.}
\label{pauli}
\end{figure}

A parity check measurement can also be measured on multiple patches (see Fig.~\ref{pauli} for an example). 
This measurement procedure is performed by first initializing an ancilla patch. The product of the operators on the boundaries of qubit patches can be measured only if they share a border with (adjacent to) the ancilla patch. For two-qubit cases, we may wish to measure any of the four possible combinations $XX$, $XZ$, $ZX$, or $ZZ$, but this is only possible if the corresponding $X$ and $Z$ operators border the ancilla patch.

After the chosen $X$ or $Z$ boundaries are merged with and then split from the ancilla patch, the ancilla patch is measured. (Combinations with $Y$ are also possible if both $X$ and $Z$ boundaries for a patch border the ancilla.)

The ancilla patch measurement projects the qubits onto the $+1$ or $-1$ eigenspace of the multi-qubit Pauli product operators and removes the ancilla patch from the board. This operation costs 1 Tock.
 
\paragraph{Patch deformation and rotation}
A patch can also be expanded to cover more tiles (1 Tock) or shrunk to fewer tiles (0 Tocks).
$X$/$Z$ corners can be moved along the boundaries of the patch (1 Tock). 
This deformation of a patch can be useful when we want to expose more operators on the patch boundary to perform multi-patch measurements or to switch between solid and dashed edges.

%% file: ourapproach.tex
\subsection{Framework Overview}

The general approach to creating graph states is to apply controlled-phase gates between qubits of adjacent vertices in the graph state~\cite{cabello_optimal_2011}, which can be performed by two two-party parity check measurements. 
The worst case, when all controlled-phase gates are performed sequentially, would require $O(|E|)$ Tocks which for a dense graph will scale on the order of $O(|V|^2)$.
For this reason, we opted for the approach of using multi-party parity check measurements to directly project the state onto the eigenspace of each of the stabilizer generators.
In contrast to the controlled-phase gate approach, this will scale only on the order of $O(|V|)$ Tocks and is more efficient than the controlled-phase gate method (also see Table.~\ref{tab:depth}).
The primary goal of this approach is to minimize and parallelize these stabilizer generator projections for optimal efficiency.

\begin{figure}[btp]
\centering
\includegraphics[width=0.29\textwidth]{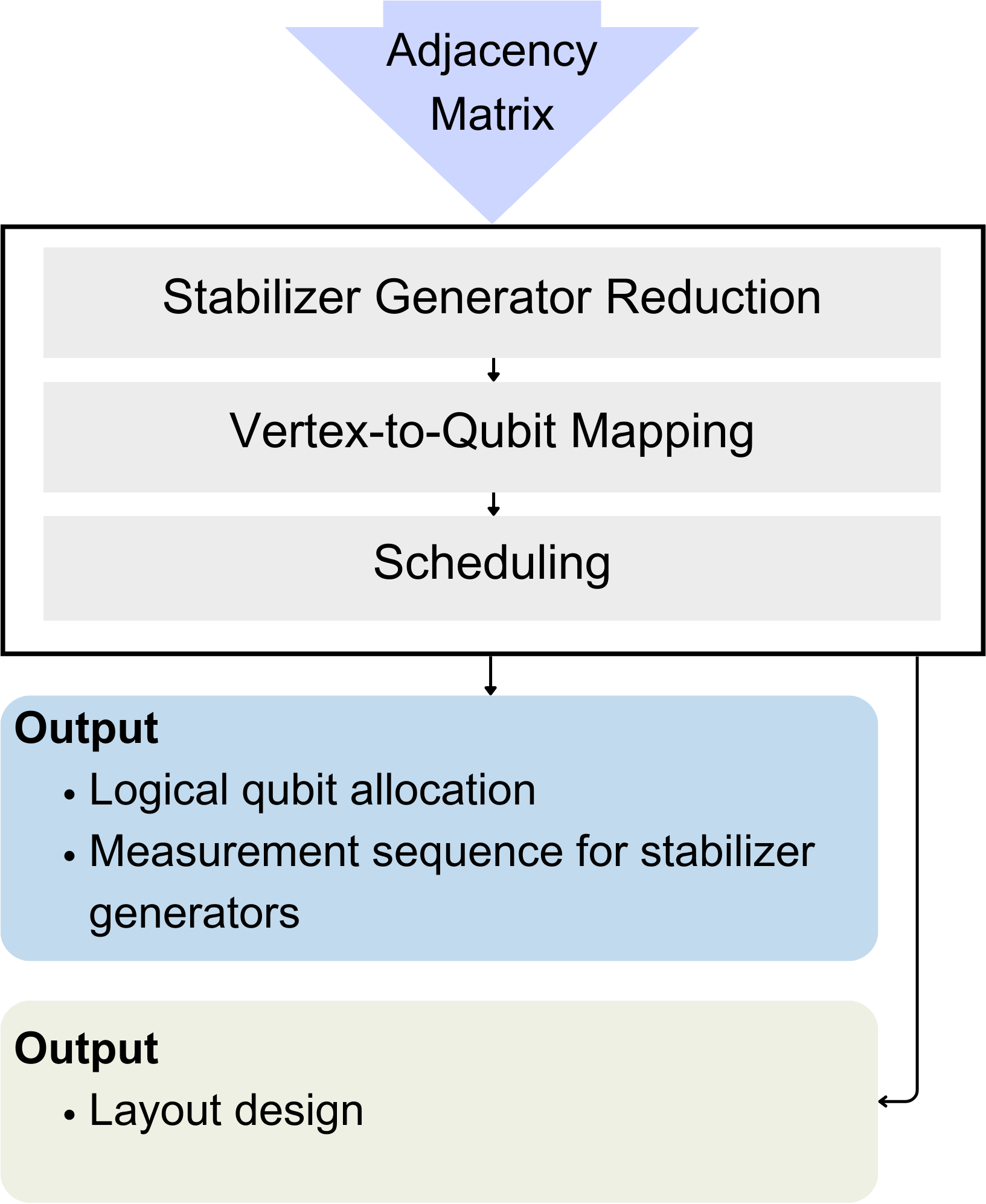}
\caption{Flow chart for Substrate Scheduler.}
\label{Fig.Frame}
\end{figure}

Fig.~\ref{Fig.Frame} summarizes the key components of the Substrate Scheduler. The Substrate Scheduler uses the adjacency matrix as input, which theoretically allows it to process any graph state that is represented as an adjacency matrix. 

The first step in the process is to reduce the number of stabilizer generators that need to be actively measured. This is achieved by appropriately initializing the qubits so that they are already stabilized by a subset of the stabilizer generators. The next step is to determine an optimized mapping of the qubits. We proposed a heuristic scheme that works well for certain types of graphs. After that, the Substrate Scheduler finds an optimized schedule for the stabilizer generators and minimizes the time cost by concurrently measuring as many stabilizers as possible. This results in the optimized schedule and a logical qubit allocation, which can be combined with the default layout design and passed on to the final compilation phase for the target fault-tolerant quantum computer to generate a fault-tolerant quantum graph state.

\subsection{Design of the Layout} 
\begin{figure}[tbp]
\centering
\includegraphics[width=0.475\textwidth]{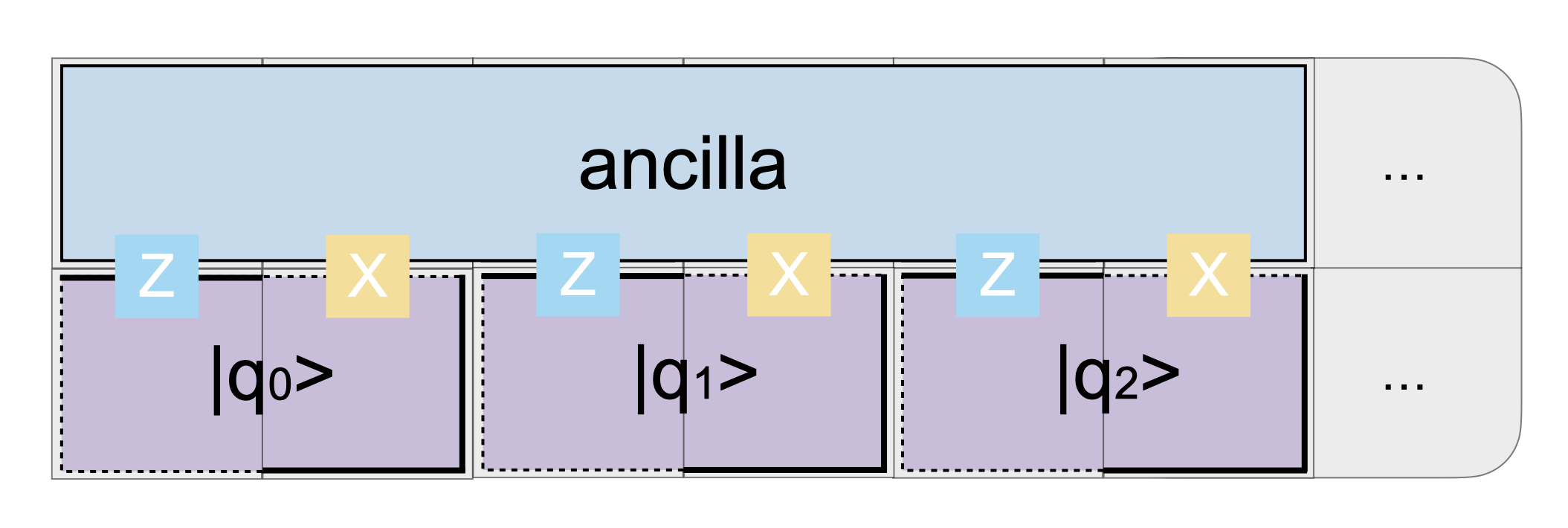}
\caption{\label{fig:2_row_layout} The 2-row layout design predominantly used in this paper has a spatial cost of $4n$ tiles, where $n$ is the number of vertices in the graph. It should be noted that this spatial cost is not optimal when utilizing the pre-mapping stabilizer generator reduction and allowing the use of one-tile one-qubit patches.}
\end{figure}

Given the space-time trade-offs,  our primitive design is to use the two-tile one-qubit patches (see Section \ref{Preliminaries}) proposed in the paper as logical qubits to represent the vertices in the graph state. This design enables us to measure stabilizer generators with the ancilla patch without requiring costly patch rotations, as demonstrated in \cite{litinskiGameSurfaceCodes2019}, which may otherwise dominate the graph state generation process. We assume that the ancilla patch consists of a row (one bus) of logical qubits of the same length as the two-tile one-qubit patches. Then the structure we use is a block of $2$ rows by $2n$ columns representing $n$ logical qubits and the corresponding ancilla patch, as shown in Fig.~\ref{fig:2_row_layout}. 
This configuration requires $4n$ tiles corresponding to $\sim 8d^2$ physical qubits (see Section~\ref{Preliminaries}).
Note that the implementation of magic state distillation is not considered in our cost calculation for the time being.

\subsection{Three-Phase Process of Optimization}
We split our process which reduces the time steps required to create a graph state via stabilizer formalism into 3 phases, namely: stabilizer generator reduction, vertex-to-qubit mapping, and scheduling of the stabilizer generator measurements. 

\subsubsection{Stabilizer Generator Reduction}
As mentioned in Section \ref{Preliminaries}, an $n$-qubit graph state $\ket{G}$ is uniquely identified as the simultaneous $+1$ eigenstate of all $n$ stabilizer operators, i.e. we need to perform $n$ stabilizer generator measurements to generate the graph state.

From the rule set in \cite{litinskiGameSurfaceCodes2019}, we know that the time cost to initialize logical qubits to $\ket{+}$ or $\ket{0}$ are both $0$ time steps.
This allows us to initialize the qubits in such a way that they are already stabilized by a subset of the stabilizer generators. For example, the three-vertex path graph $G$ is stabilized by the following generators,
 \begin{equation*}
\begin{aligned}
g_0 &= X \otimes Z \otimes I, \\
g_1 &= Z \otimes X \otimes Z, \\
g_2 &= I \otimes Z \otimes X.
\end{aligned}
\end{equation*}
Initializing $q_0$ in $\ket{+}$ and $q_1$ in $\ket{0}$ prepares a state that is automatically stabilized by $g_0$ (here we assume that the mapping of the vertex-to-qubit is also sequential from left to right).
We can go further and initialize qubit $q_2$ in the state $\ket{+}$, which will ensure that the three-qubit state is a simultaneous +1 eigenstate of both $g_0$ and $g_2$,
\begin{align}
    g_0 \ket{+0+} = \ket{+0+} = g_2 \ket{+0+}.
\end{align}
This results in the reduction of total stabilizer generator measurements that we need to perform via multi-Pauli product measurement.
In our example, the only stabilizer generator that remains to be measured in order to prepare the three-qubit graph state is $g_1$.

\begin{figure}[btp]
\centering
\includegraphics[width=0.475\textwidth]{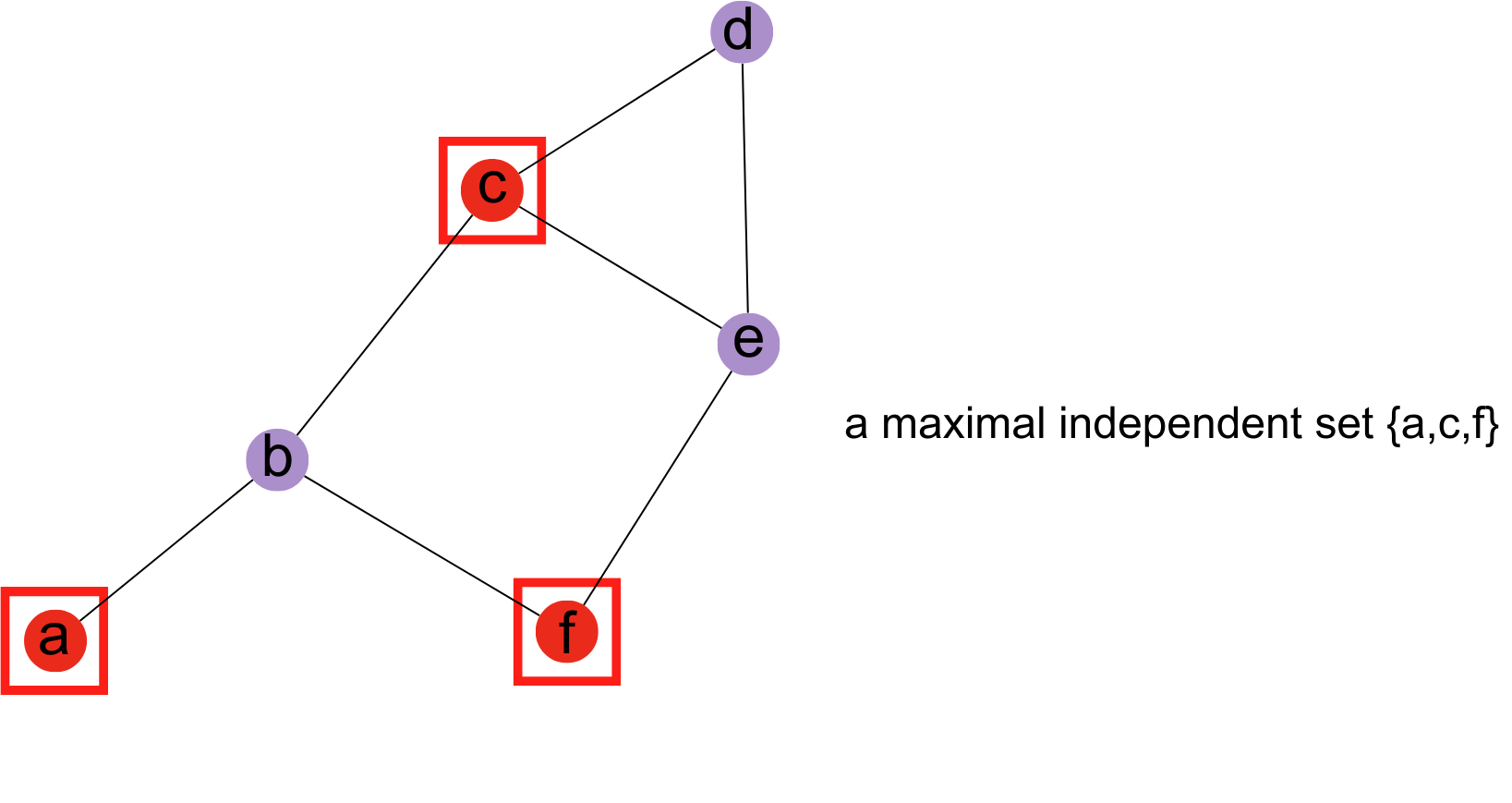}
\caption{An example of a maximal independent set. By initializing the logical qubits corresponding to the vertice $a$ in $\ket{+}$, the requirement to measure the stabilizer generator $g_a$ associated with this vertex can be omitted.}
\label{max}
\end{figure}
 
For a simple argument, to stabilize $g_a$, we must initialize qubit $a$ in the state $\ket{+}$, and all of its neighboring qubits in $\ket{0}$. This initialization strategy prohibits us from simultaneously stabilizing any of the neighboring qubits in $\operatorname{ngbr}(a)$. However, we can stabilize qubits inside $\operatorname{ngbr}(\operatorname{ngbr}(a))$ that are not adjacent to $a$ concurrently with $g_a$.

As a result, the optimal reduction problem can be reduced to the maximum independent set problem.
Knowing the maximum independent set $\alpha(G)$ identifies the qubits which must be initialized in $\ket{+}$, while all the remaining qubits are initialized in $\ket{0}$.
This means that the number of stabilizer generators that need to be measured decreases to $| V | - |\alpha(G)|$.
In practice, we compute a maximal independent set (see Fig.~\ref{max}) using the greedy algorithm in NetworkX \cite{hagberg2008exploring} to achieve a reasonable time complexity.
 
\subsubsection{Scheduling of the Stabilizer Generator Measurements}

As shown in Fig.~\ref{Fig.Frame}, the mapping of qubits to vertices is conducted before the scheduling of the stabilizers, but here we find it convenient to motivate the mapping problem by explaining the scheduling problem first.

\begin{figure}[tbp]
\centering
\includegraphics[width=0.485\textwidth]{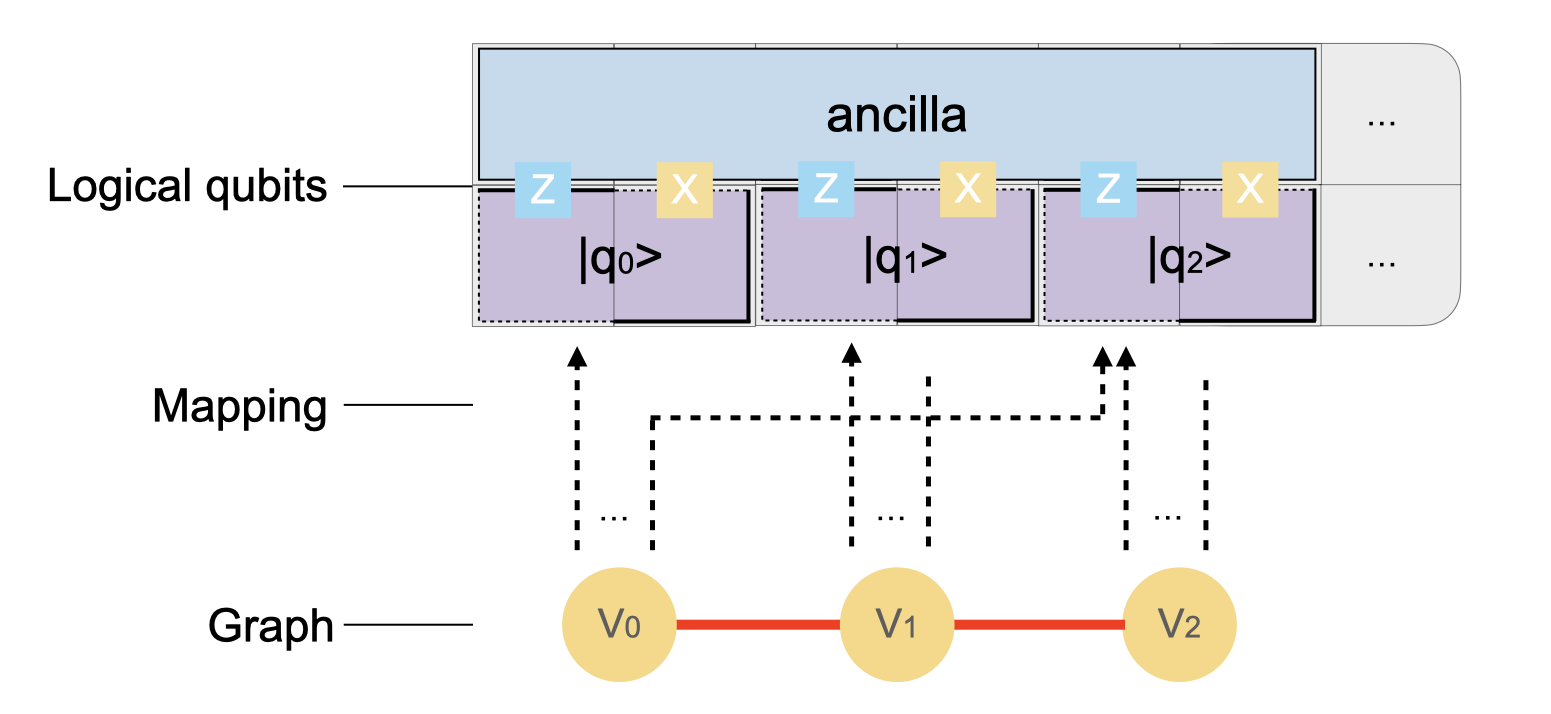}
\caption{A schematic diagram of the mapping process. To map the vertices in a graph to logical qubits, one can assign labels to each logical qubit.}
\label{map}
\end{figure}

Given that each vertex in the graph is assigned to a specific logical qubit as depicted in Fig.~\ref{map}, along with a fixed stabilizer generator reduction, we can decide the measurement sequence for the stabilizer generators by exploiting their commutativity. 
To measure a stabilizer generator, we need an ancilla that covers all the mapped vertices of the stabilizer.
This implies that we cannot measure any two stabilizer generators whose mapped vertices overlap between the leftmost and rightmost qubits at the same time.
In order to measure a stabilizer generator $g_a$, the ancilla is only required to cover patches that represent vertices $a$ and $\operatorname{ngbr(a)}$, without the need to cover the remaining qubits.
Thus, we can define an ancilla block for measuring the generator $g_i$ by a pair of two numbers $(L_i, R_i)$ with $L_i, R_i \in [1, 2n]$ where $L_i$ and $R_i$ denotes the leftmost and the rightmost qubits that the ancilla needs to cover, respectively.

The problem of maximizing the number of stabilizer generators that can be measured simultaneously in a single step can be rephrased as the problem of minimizing the total height of the stacked ancilla blocks.
An optimal solution to stabilizer measurement scheduling can be found with a simple greedy algorithm where we first sort the list of pairs $[(L_i, R_i)]$ in non-decreasing order of $R_i$. If two pairs have the same $R_i$ value, we sort them in non-decreasing order of $L_i$ as well.
To find the optimal solution, for each time step, we traverse the sorted list from the beginning while maintaining a set $set_t$ of generators that we will concurrently measure. For each pair, if the pair $(L_i, R_i)$ has $R_j < L_i$ for all $j$ in $set_t$, we can remove it from the unmeasured list and add it to $set_t$ and proceed with traversing the list until the end of the list.
We then repeat this until the sorted list is empty, and the $set_t$ we get at each iteration corresponds to the stabilizer measurements we can perform simultaneously in each time step.

We can see that this approach is optimal by simple contradiction argument.
Suppose our approach is not optimal: when considering $(L_i, R_i)$, we should skip it and we could pack more blocks, namely $(L_j, R_j)$ and $(L_k, R_k)$ into the set where $R_i \leq R_j \leq R_k$.
However, since we traverse the list in non-decreasing order of $R$, if $(L_j, R_j)$ and $(L_k, R_k)$ can be concurrently measured, $(L_i, R_i)$ and $(L_k, R_k)$ can also be concurrently measured, which gives a contradiction.
This contradiction proves the optimality of our approach.
This scheduling algorithm on average can be done in $O(n \log n)$ time and $O(n)$ space.

\subsubsection{Vertex-to-Qubit Mapping}
Qubit mapping (i.e. mapping vertices to logical qubits) affects the number of stabilizers that can be measured simultaneously. We want to rearrange the vertex to patch assignment in order to achieve a higher degree of parallelism when later scheduling stabilizers such that the resulting time step after scheduling is minimized. 

In mapping, we aim to minimize the overlap of stabilizer generators (or ancilla) to maximize the potential for simultaneous multi-Pauli measurements. 
Intuitively, we position the dense components of the graph $G$ in adjacent positions as much as possible to minimize the distance between the start and end positions $(L, R)$ of the stabilizer generators. 
We use an iterative mapping method that involves finding the minimum cut of the graph (see Appendix \ref{Appendix}). In graph theory, a minimum cut refers to the partition of the vertices of graph $G$ into two sets that minimize the number of edges crossing the partition. For the MinCut mapper, we repeat this process of cutting the graph (the first subgraph) until we obtain a subgraph with two or fewer vertices. We then map these vertices to adjacent logical qubits at the end of the row and remove this subgraph. Then, we continue processing the first subgraph until there are no subgraphs with more than two vertices remaining.

An efficient randomized algorithm, Karger's algorithm~\cite{10.5555/313559.313605}, can be used to solve the minimum cut problem. 
The time complexity of a single run is $O(n^{2})$, To obtain the optimal solution, we can run the algorithm $n^{2}\log(n)$ times, resulting in an overall time complexity of $O(n^{4}\log(n))$. 

\begin{figure}[bp]
\centering
\includegraphics[width=0.475\textwidth]{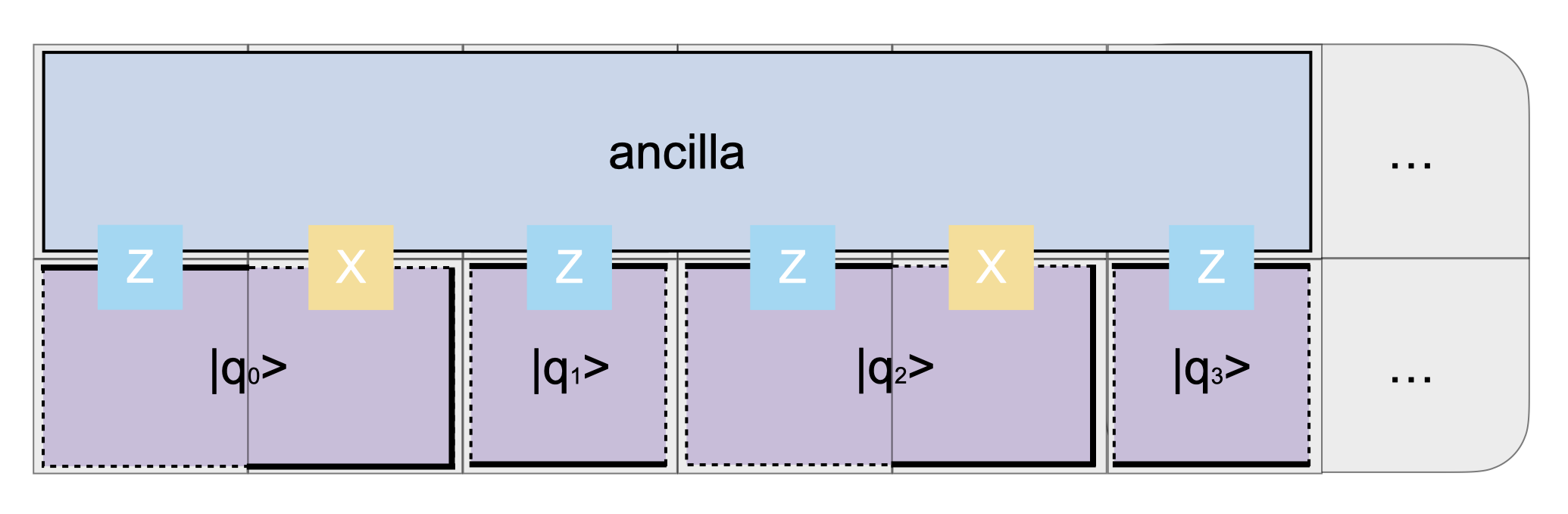}
\caption{The 2-row layout design after the pre-mapping stabilizer generator reduction. The logical qubits are depicted by both two-tile one-qubit patches and one-tile one-qubit patches. \label{fig:improved_layout}}
\end{figure}

Further optimization can be performed after the initial stabilizer generator reduction and the vertex-to-qubit mapping.
Any qubit $a\in\alpha(G)$ is initialized in the $\ket{+}$ state, which means that it will not take part in stabilizer measurements where it needs to be measured in the $X$ basis.
This allows us to represent such qubits by patches that take up a single tile, as pictured in Fig.~\ref{fig:improved_layout}.
The space complexity required after this optimization is reduced to $2\times 2n-|\alpha(G)|$.

%% file: evaluation.tex
We conducted tests on the functionality and performance of the Substrate Scheduler. Since in the current version the layout has been fixed and the space cost is thus determined, our focus in this work is on evaluating the time cost of creating the graph state. In the stabilizer formalism, creating a graph state with $n$ vertices involves measuring $n$ stabilizer generators. Without any optimization, this would require $n$ time steps in the worst case as measuring a stabilizer generator takes $1$ time step. All time steps mentioned in this section refer to the time measure known as ``Tock'', which was introduced in ~\cite{litinskiGameSurfaceCodes2019} (see Section~\ref{Preliminaries}).

To assess the performance of our method, we will compare the time cost of creating the graph state after applying the Substrate Scheduler to the initial number of time steps $n$.
Describing algorithm-specific graph states corresponding to different quantum circuits or algorithms can be challenging due to their distinct characteristics, such as size and density, especially after undergoing optimizations like local complementation \cite{hein2004multiparty, PhysRevA.69.022316}.
It is not yet clear what structural properties we should expect from a typical instance of an algorithm-specific graph state.

In order to gain insight into the performance of the Substrate Scheduler, we chose the following testing strategy.
We begin by testing specific regular types of graphs to verify that Substrate Scheduler produces the correct and expected output.
Then we move on to testing how the sparsity and size of the graph affect the total reduction in time steps needed to prepare the target graph state.

\subsection{Evaluation of Specific Types of Graphs}

We begin by evaluating the Substrate Scheduler's performance on common classes of graphs, including path graphs, star graphs, tree graphs, and complete graphs.
These classes of graphs were chosen because they allow for analytical optimization of the graph state preparation procedure, enabling us to verify the correctness of the Substrate Scheduler's output.
Understanding how our approach to reducing the overhead of graph state preparation behaves in these simple examples also provides useful intuition about what to expect when we test the Substrate Scheduler on random graphs.

    \begin{figure}[tb]
        \centering
        \begin{subfigure}[b]{0.24\textwidth}
            \centering
            \includegraphics[width=\textwidth]{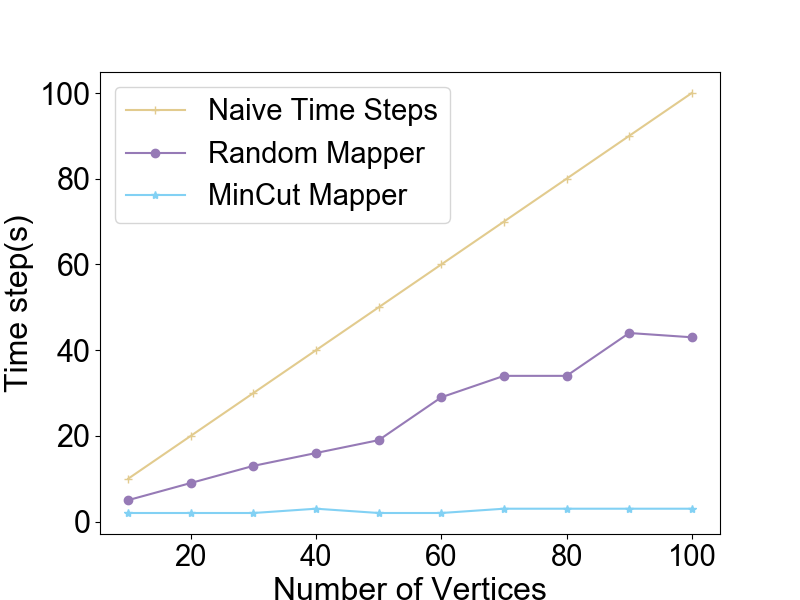}
            \caption[Path Graph]%
            {{\small Path Graph}}    
            \label{fig:line}
        \end{subfigure}
        \hfill
        \begin{subfigure}[b]{0.24\textwidth}  
            \centering 
            \includegraphics[width=\textwidth]{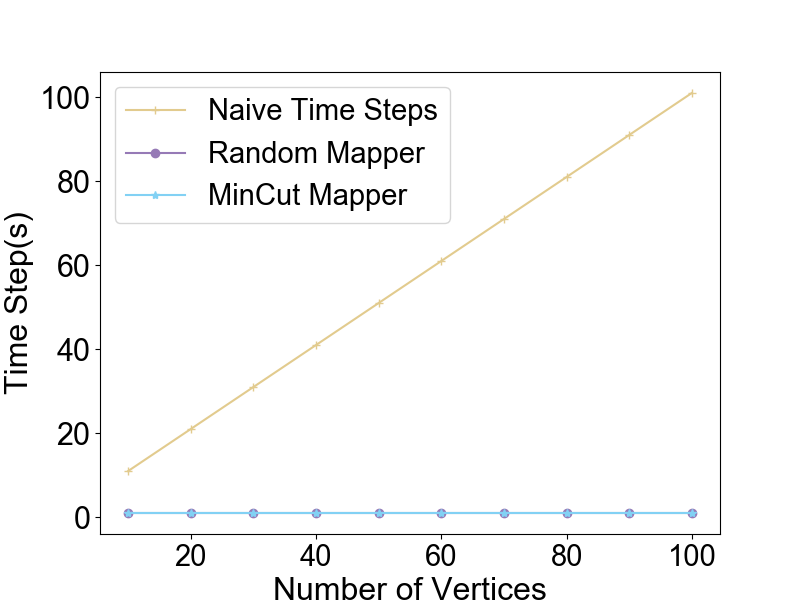}
            \caption[]%
            {{\small Star Graph}}    
            \label{fig:star}
        \end{subfigure}
        \vskip\baselineskip
        \begin{subfigure}[b]{0.24\textwidth}   
            \centering 
            \includegraphics[width=\textwidth]{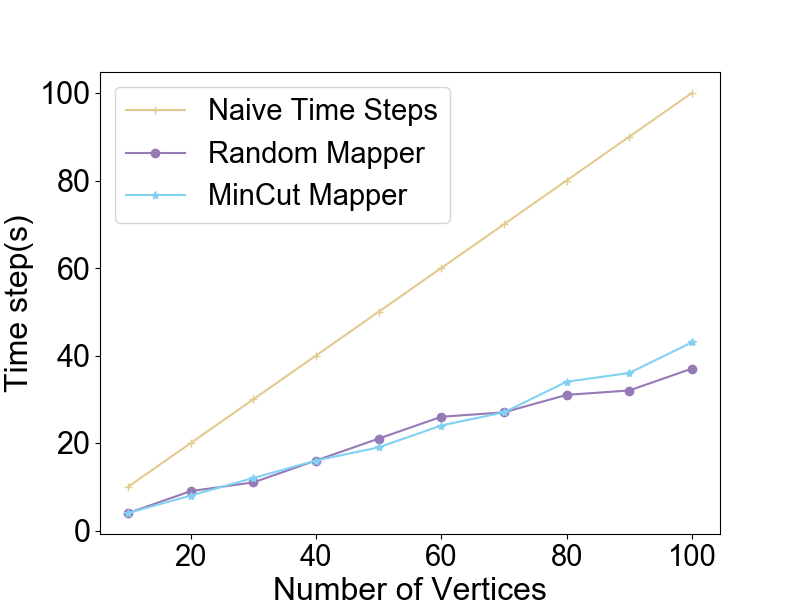}
            \caption[]%
            {{\small Tree Graph}}    
            \label{fig:tree}
        \end{subfigure}
        \hfill
        \begin{subfigure}[b]{0.24\textwidth}   
            \centering 
            \includegraphics[width=\textwidth]{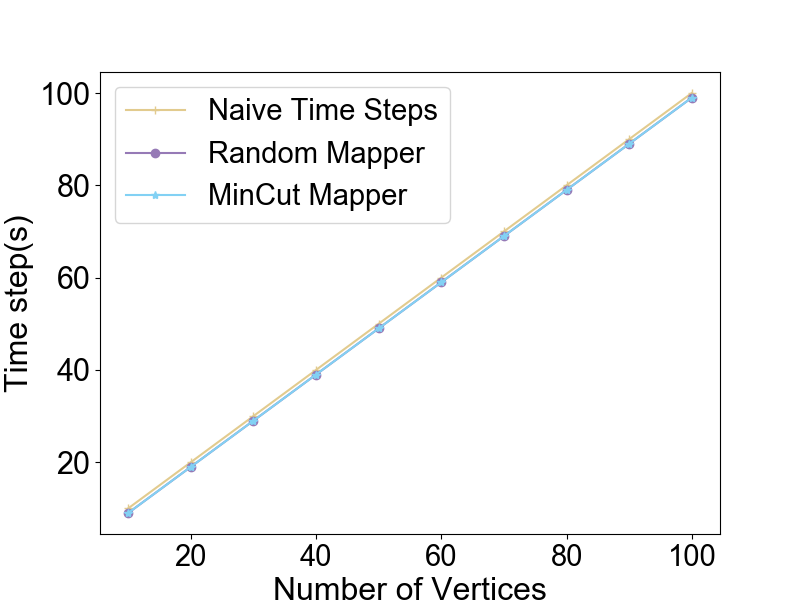}
            \caption[]%
            {{\small Complete Graph}}    
            \label{fig:complete}
        \end{subfigure}
        \caption
        {The performance of the Substrate Scheduler on four distinct types of graphs. The horizontal axis indicates the graph size (number of vertices), while the vertical axis shows the number of time steps required to complete the initialization of the graph state. Light blue lines represent experiments conducted with all optimization techniques, including the proposed MinCut mapping method, while grey lines depict experiments conducted with a random mapping method (no mapping optimization). Overall, the Substrate Scheduler achieves a significant speedup in terms of time steps across three out of four graph types, with path graphs and star graphs experiencing a reduction in time steps from linear growth to constant levels. In the complete graph, there is no discernible improvement. Notably, for path graphs, the proposed MinCut mapping method in this study outperforms random mapping, while for the other three types of graphs, the MinCut mapping method achieves performance that is not inferior to random mapping.} 
        \label{fig:4kinds}
    \end{figure}

Fig.~\ref{fig:4kinds} illustrates the performance of the Substrate Scheduler when processing these specific types of graphs.
We observe that output reproduces the expected results.
The analytic optimization is summarized in Table~\ref{tab:depth}.
In order to test the performance of our chosen qubit mapping algorithm, we compared how the Substrate Scheduler performs when we use the MinCut mapping and a random qubit assignment.
For the path graph, the qubit assignment is crucial as observed in Fig.~\ref{fig:4kinds}(a).
The MinCut mapping clearly outperforms a random qubit assignment. 

\begin{table*} [h]
  \setlength{\tabcolsep}{0.7\tabcolsep}
  \centering
  \begin{tabular}{ *{4}{c} }
    \toprule
    \textbf{Graph Type} & \textbf{CZ Preparation Depth} &
    \textbf{Maximum Stabilizer Reduction} & 
    \textbf{Parity Check Preparation Depth} \\
    \midrule
    Path Graph  & $2$ & $|V| \rightarrow |V|/2$ & 2\\
    Star Graph    & $|V|-1$ & $|V| \rightarrow 1$ & 1\\
    Random Tree  & $\Delta(G)$ & $|V| \rightarrow |V|/2$ & $\Delta(G)$ \\
    Complete Graph & $|V|$ or $|V|-1$ & $|V| \rightarrow |V|-1$ & $|V|-1$\\
    \bottomrule
  \end{tabular}
\caption{Costs associated with generating different types of graph states, where $\Delta(G)$ is the highest degree of any vertex in $G$. The second column is the optimal preparation depth for preparing the graph state with CZ gates, which means the fewest number of time steps to prepare the graph state when ancilla buses are not restricted by the spatial layout, proportional to the chromatic index of the graph. The third column is the number of parity checks that must be performed before and after the maximum independent set of this graph class is initialized in $\ket{+}$ for the stabilizer generator reduction. The fourth column is the smallest amount of time steps taken to perform the stabilizer parity checks when allowing non-overlapping parity check to be applied in parallel. The naive time step cost for generating graph states by stabilizer formalism for any graph is $|V|$, (corresponding to the number of stabilizer generators that need to be applied), and the complete graph has preparation depth $|V|$ when $|V|$ is odd, and $|V|-1$ when $|V|$ is even. The random tree maximum stabilizer reduction is a lower bound, achievable when ancilla buses are not restricted by the spatial layout, since trees are bipartite and we choose the maximal independent set to be the larger of the two bipartitions.}
\label{tab:depth}
\end{table*}

These numbers can be compared to the theoretical minimum, assuming no resource constraints such as the availability of the ancilla bus. When unconstrained by resources, the time steps taken to prepare a graph state is either the maximum degree of $G$, written $\Delta(G)$, or $\Delta(G)+1$\cite{cabello_optimal_2011}. This is because of Vizing's theorem \cite{vizing1964estimate}, which bounds the chromatic index into either $\Delta(G)$ or $\Delta(G)+1$. Trees always fall into the first class of requiring only $\Delta(G)$ time steps \cite{fournier1973colorations}. Table~\ref{tab:depth} further discusses the optimal time step for each type of graph, including the theoretically optimal time step obtained after applying all three optimization methods.

The experimental results show that when using the MinCut mapper, the path graph, the star graph, and the complete graph all achieved their theoretical optimal values, whereas Substrate Scheduler cannot further optimize the complete graph. The time step growth of the random tree depends on the maximum dimension of the graph, as previously discussed.
    
\subsection{Evaluation of Random Graphs with Different Density}

\begin{figure}[H]
\includegraphics[width=0.5\textwidth]{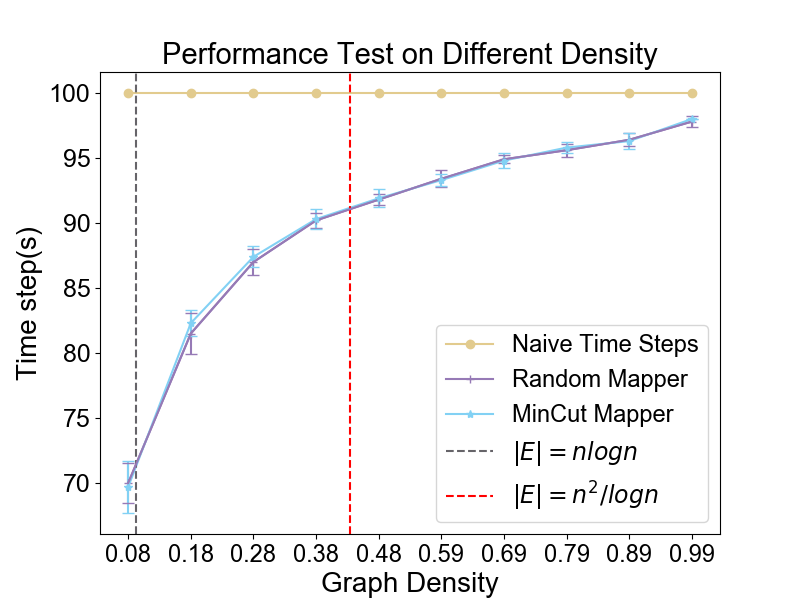}
\caption{The performance of the Substrate Scheduler is analyzed across a range of densities. Each point on the graph represents a random connected graph with 100 vertices.
The horizontal axis indicates the density of the graph. 
To the left of the gray dashed line, graphs are sparse; between the gray and red lines, graphs are intermediate; and to the right of the red dashed line, graphs are dense. Each point on the plot represents the average performance over $10$ randomly generated instances, and the error bars indicating the standard deviation. The results demonstrate the reduction in time cost achieved by the Substrate Scheduler, which becomes less significant as graph density increases. Notably, the time cost growth approximately follows a logarithmic curve, initially increasing rapidly before gradually slowing down as the graph changes from sparse to dense. Additionally, the MinCut mapping method exhibited performance comparable to that of the random mapping method. 
}
\label{Fig.connectivity}
\end{figure}

The type of graph, its size, and its density are all significant factors that contribute to the overall time cost. The experiments in Fig.~\ref{Fig.connectivity} show the effect of different graph densities on the time cost.

The graphs in the experiment are generated by the Python package NetworkX~\cite{hagberg2008exploring} by uniformly selecting from the set of all graphs containing 100 vertices and $|E|$ edges. The density of graphs is represented by the ratio between the number of edges in a graph $|E|$ and the maximum number of edges that the graph can contain. Thus when the number of vertices is fixed, the number of edges has determined the density of a graph. For undirected simple graphs, we define the graph density as:
\begin{equation}
   \operatorname{Density}=\frac{2|E|}{|V|\times(|V|-1)}
\end{equation}

The result shows that it performs better on sparse graphs compared to dense graphs. However, with increasing graph density, the growth of time cost after optimization approximately follows a logarithmic curve. This suggests that the impact on the Substrate Scheduler's performance becomes less significant as the density of the graph reaches a certain level.

\subsection{Scalability Testing}

\begin{figure*}[h]
        \centering
        \begin{subfigure}[b]{0.475\textwidth}
            \centering
            \includegraphics[width=\textwidth]{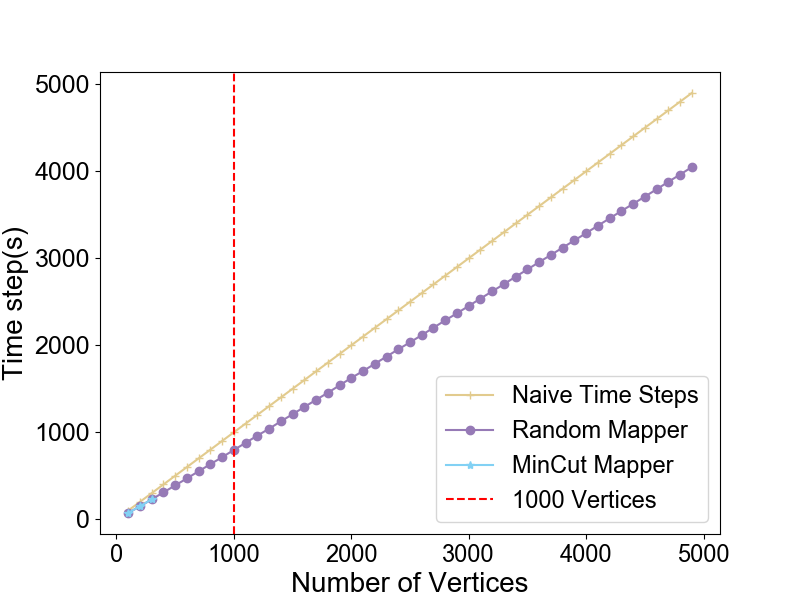}
            \caption[Path Graph]%
            {{\small Sparse graph with $n\log n$ edges}}    
            \label{fig:mean and std of net14}
        \end{subfigure}
        \hfill
        \begin{subfigure}[b]{0.475\textwidth}  
            \centering 
            \includegraphics[width=\textwidth]{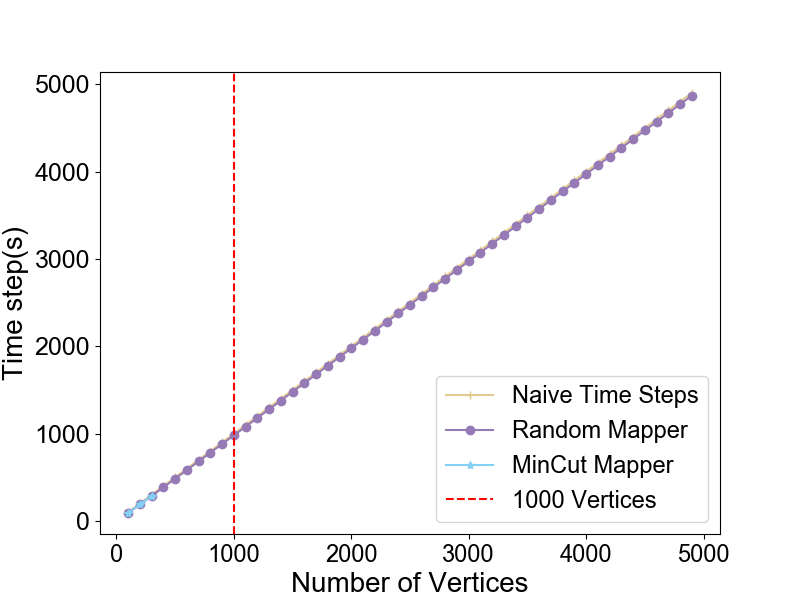}
            \caption[]%
            {{\small Dense graph with $n^{2}/\log n$ edges}}    
            \label{fig:mean and std of net24}
        \end{subfigure}
        \vskip\baselineskip
        \caption
        {Performance of Substrate Scheduler on graphs of varying sizes. To ensure representative results, we selected two types of graphs of different densities, as defined in this work: sparse graphs (with $O(n\log n)$ edges) and dense graphs (with $O(n^{2}/ \log n)$ edges) 
        Each point on the plot is averaged over 10 randomly generated instances. Error bars are too small to be seen.
        The horizontal axis represents the number of vertices in the graph, while the vertical axis indicates the number of time steps.  To avoid excessive compilation times, which would become impractical in real applications, we tested with the MinCut Mapper for only up to 300 vertices. 
        The red vertical dashed line indicates the target for this study, which is the graph with 1000 vertices.
        }
        \label{fig:mean and std of nets}
    \end{figure*}

To assess the Substrate Scheduler's potential for use in promising near-term applications, we conducted experiments to evaluate its performance across graphs of different sizes (see Fig.~\ref{fig:mean and std of nets}). 
In the current testing scale, we found that the performance of MinCut mapper and random mapper was comparable.

Our experiments showed that the Substrate Scheduler can effectively handle sparse graphs, and thus we can expect its application to larger graphs in the future.

For dense graphs, we did not observe any significant reduction in the number of time steps.
This is not a serious limitation as dense graphs can be transformed into sparse ones without affecting the quantum computation itself, as we discuss in Section~\ref{sec:discussion}.
In our study, we set our target at generating  1000-vertex graph states, which provides a reasonable evaluation of the tool's general behavior and compatibility with other compiler tools.

%% file: discussion.tex
\label{sec:discussion}

To the best of our knowledge, this work is one of the first studies for this problem, presenting a feasible approach to optimize the time step cost of generating fault-tolerant graph states.
The shape of the 2-row layout used in our work was chosen for ease of use but is not strictly required.
The ancilla bus can be easily deformed to better suit the physical layout of the chip without affecting our analysis.
On the other hand, the mapping approach chosen is optimal only for specific graph types.

Our scalability testing shows comfortable scaling up to 1,000 nodes in the graph.  The graph size needed for application circuits will depend on the number of input qubits and the number of T gates in the circuit, which in turn is driven by the required precision for the Solovay-Kitaev decomposition~\cite{Kitaev_1997, selinger2014efficient, 10.5555/2011679.2011685} for many algorithms. Graphs far larger than 1,000 nodes will be needed, but taking into consideration various optimizations that are being concurrently developed, it is difficult to estimate sizes at the time of this work, so we defer that estimation to future work.

The density of graphs that will arise in the production use of the toolchain is not yet well understood. The testing presented here focuses on several abstract types of graphs, both because of the difficulty of predicting the graph structures (which will also change as a result of the future work described below) and in order to provide a solid basis for understanding the performance and for debugging and algorithm development.

The following are some extensions to this work that could both improve our implementation performance and lead to more rigorous optimality bounds:
\paragraph{Local complementation}
Local complementation (LC) is an operation that can be used to transform graph states, via single qubit gates, into a large class of equivalent graph states with highly varied structures~\cite{adcock_mapping_2020}. A potential optimization of our method is to reduce the preparation depth of the algorithm-specific graph state by modifying the input state with LC; although computing the local minimum degree of a graph is both NP-complete and hard to approximate~\cite{cattaneo_minimum_2015}, LC can be used to minimize over other metrics such as graph size to reduce the cost of graph state constructions~\cite{cabello_optimal_2011}. For example, the complete graph can be transformed into a star graph to reduce the time steps for its preparation.

After the stabilizer generator reduction is performed on an LC-optimized graph state, further optimization is possible by using LC to then reduce the connectivity constraints required by the stabilizer checks. It is presently unknown if these two LC optimization steps would meaningfully differ. 
\paragraph{Stabilizer generator reduction}
In our approach, we choose to maximize the number of stabilizer reductions by approximating the maximum independent set of the input graph state, with a maximal independent set. This will minimize the number of stabilizer checks required, however, this does not necessarily minimize the preparation depth required to perform the stabilizer checks; it is plausible that for some algorithm-specific graph states, there is a stabilizer generator reduction which contains more stabilizer checks but can be performed in parallel with fewer time steps. In future work, it may be possible to choose the stabilizer generator reduction according to different criteria to more rigorously guarantee preparation depth optimality bounds.
\paragraph{Optimizing mapping methods}
It is not known currently if there is an asymptotic polynomial time algorithm for vertex-to-qubit mapping which minimizes preparation depth, even for the linear architecture proposed. The preparation depth overhead in this step is due to the geometric properties of the proposed linear architecture which limits which parity checks can be done simultaneously. It is not known to what extent this problem is intractable for more structurally complex ancilla bus architectures. 

\paragraph{Other ancilla bus architectures}
The linear ancilla bus architecture is promising because it has clear constraints on geometric connectivity, which translate to a precise problem definition for optimizing the vertex-to-qubit mappings, and for having a high data to ancilla qubit ratio of 50\%. While other architectures may have a worse data to ancilla qubit ratio, they may also be able to prepare graph states with fewer time steps due to different connectivity rules about which pairs of vertices can participate in stabilizer checks simultaneously. Further research is needed to quantitatively compare different architectures in both the data to ancilla qubit ratio and the overall preparation depth reduction, to see which architectures can perform with the lowest overall time-space volume.

%% file: Appendix.tex
The pseudocode for the MinCut mapper (see Section \ref{Approach}) is provided below as Algorithm~\ref{psedo}.
 \begin{algorithm} [h]
 \caption{Algorithm for MinCut Mapper}
 \begin{algorithmic}[H]
\algnewcommand{\LeftComment}[1]{\Statex \(\triangleright\) #1}
 \renewcommand{\algorithmicrequire}{\textbf{Input:}}
 \renewcommand{\algorithmicensure}{\textbf{Output:}}
 \Require $g$: the input graph
 \Ensure  $mapping$: the indexes of the vertices that have been mapped to the logical qubits
 \State $mapping \gets Array[]$ \Comment{Initialize the mapping array}
 \State $G\gets deepcopy(g)$ \Comment{Create a copy of the input graph}
 \Function{min\_cut}{$graph,component$}
 \State $cut\_edges\_list \gets Array[]$ \Comment{Initialize the list of cut edges}
      \State $shortestLen = very\_large\_value$
        \State $shortestV \gets Array[]$

\For {$i = 0$ to $num\_of\_repetitions$} \Comment{Loop over iterations}
\State $currentLen, currentV \gets karger(graph)$ \Comment{Apply Karger's algorithm to find the minimum cut. The function $karger$ takes a graph as input and returns two values: the number of edges that need to be cut $currentLen$, and the vertices of those edges $currentV$.}
  \If {$currentLen<shortestLen$}
  \State $shortestLen\gets currentLen$
  \State $shortestV \gets deepcopy(currentV)$
  \EndIf
  \EndFor
  \For {$edges$ in $currentV$}
  \State {$cut\_edges\_list.append(edges)$}
\EndFor
 \State $graph.remove\_edges\_from(cut\_edge)$ \Comment{Remove the cut edges from the graph}
 \State \Return $graph$ 
 
 \EndFunction
 \Function{mapping\_min}{$G$}

  \For {$component$ in $connected\_subgraphs(G)$}
  \If {$(number\_of\_vertices\left(component\right)>2)$}
  \State $G \gets min\_cut(G,component)$
  \State \Return $mapping\_min(G)$ \Comment{Recursively call the function on the updated graph}
  \Else
  \State $mapping. append (component. vertices)$
  \State $G.remove\_vertices\_from(component.vertices)$
  \State\Return $mapping\_min(G)$ \Comment{Recursively call the function on the updated graph}
  \EndIf
  \EndFor
\EndFunction
\State $mapping\_min(G)$
\State \Return $mapping$ 
 \end{algorithmic} 
 \label{psedo}
 \end{algorithm}